\documentstyle[aps,eqsecnum,prb,floats]{revtex}

\begin{document}
\draft
\twocolumn
\input epsf

\title{LEED holography applied to a complex superstructure:\\
a direct view of the adatom cluster on SiC(111)-(3$\times$3)}

\author{K. Reuter, J. Schardt, J. Bernhardt, H. Wedler, U. Starke and K. Heinz}
\address{Lehrstuhl f\"ur Festk\"orperphysik, Universit\"at Erlangen-N\"urnberg,
Staudtstr. 7, D-91058 Erlangen, Germany\\
\smallskip
\small
\begin{quote}
For the example of the SiC(111)-(3$\times$3) reconstruction we show that
a holographic interpretation of discrete Low Energy Electron
Diffraction (LEED) spot intensities arising from ordered, large unit cell
superstructures can give direct access to the local geometry of a cluster 
around an elevated adatom, provided there is only one such prominent atom per
surface unit cell. 
By comparing the holographic images
obtained from experimental and calculated data we illuminate validity, current 
limits and possible shortcomings of the method. In particular, we show
that periodic vacancies such as cornerholes may inhibit the correct detection
of the atomic positions. By contrast, the extra diffraction intensity 
due to slight substrate reconstructions, as for example
buckling, seems to have negligible influence on the images. 
Due to the spatial information depth of the method the
stacking of the cluster can be imaged down to the fourth layer. 
Finally, it is
demonstrated how this structural knowledge of the adcluster geometry can be
used to guide the dynamical intensity analysis subsequent to the holographic
reconstruction and necessary to retrieve the full unit cell structure.
\flushleft{PACS-numbers: 61.14.Nm, 61.10.Dp, 61.14.Hg, 68.35.Bs \hfill
{\bf Phys.\,Rev.\,B, (1998) {\em in press}}
}
\end{quote}
}
\maketitle
\narrowtext

\section{Introduction}
The majority of ordered surface structures known today
has been determined by Low Energy Electron Diffraction (LEED)
\cite{NIST}. The strong elastic and inelastic interaction
of electrons in the energy range 50-500eV involves a particular
sensitivity to the atomic arrangement within the outermost layers.
In many cases this permits a structure determination with a precision
of a few hundredths of an {\AA}ngstr{\o}m, making LEED one of
the primary surface crystallography methods. 
Unfortunately, the strong multiple scattering implied by this 
type of interaction also tremendously complicates the theoretical 
analysis of the acquired data. In addition, the real space 
geometry can not be drawn from the intensities directly, 
so that standard quantitative LEED structure determinations have 
to apply a trial-and-error method which is frequently supported 
by structural search procedures.
Calculated diffraction intensities of a multitude of models have 
to be compared with the experimental data until eventually
a sufficiently high agreement between both is achieved
\cite{Pendry74,VanHove79,Heinz95}.

Though more and more advanced experimental and theoretical
developments have recently given access to rather
complex surface structures \cite{Heinz95}, it is just this complexity
which at a certain degree inhibits the successful application of
quantitative LEED. The number of models as resulting from the mere
combination of all coordinates of the many atoms in a large unit cell
structure becomes so huge that it is difficult if not impossible
to handle. This applies even when for example using 
automated search algorithms in multi-parameter space
\cite{Rous93} as the latter has to include the correct model. 
However, for a large unit cell structure our structural imagination is 
frequently unable to even define the type of the correct model or
the relevant part of the parameter space containing the real
structure, in which a search could then be started. Also, methods
developed in LEED to determine the atomic positions directly
still rely on an initial good guess of the real structure
\cite{Pendry88}.

The holographic approach represents a revival of the hope for
the {\em direct} disclosure of structural information. The
idea, first developed for the related Photoelectron Diffraction
\cite{Szoke86,Barton88}, aims at the determination of at least
partial features of the structure when in addition to the multiple 
scattering problem the complexity of the surface prohibits its
full retrieval. In the present case, the information provided 
consists of the local environment around an elevated atom in the
cell, which might be an adsorbate or an intrinsic adatom resulting 
from surface restructuring. Even though only some atomic positions 
at a rather coarse resolution are determined in this manner, the 
necessary consistency of the obtained structural unit with the
complete surface geometry may rule out many models directly.
Hence, the remaining parameter space can be reduced to an extent 
sufficient to allow the application of conventional surface 
crystallography methods.

The first translation of holographic schemes to the field of LEED as
proposed by Saldin and De Andres \cite{Saldin90} was restricted to
surfaces on which atoms or molecules are adsorbed in lattice gas
disorder. The lacking periodicity creates diffraction intensity also
outside the sharp substrate Bragg spots and causes a diffuse intensity
distribution on the screen (for a recent review on Diffuse LEED (DLEED)
see e.g. ref. \onlinecite{Starke96}). This appeared as the natural input
for the Fourier-like integral transform typical for holographic techniques. 
In the course of subsequent theoretical improvements a proper
reconstruction algorithm could be established that allowed to
circumvent several problems complicating the holographic interpretation of 
LEED intensities (see section II). In the present investigation we use
the latest stage of this development which allows to construct a reliable 
image of the complete 3D atomic surrounding of the elevated atom 
from data of normal incidence alone \cite{Saldin95,Saldin96,Saldin97}.

However, these theoretical achievements were based on the use of diffuse
intensity distributions emerging from disordered systems while the majority of
interesting surface structures are ordered phases, often with large superstructure
unit cells. Certainly, it would be very advantageous to obtain a partial, but 
{\em direct} information by holographic means for these {\em ordered} phases, too. 
As we briefly demonstrated recently \cite{Reuter97_2}, the diffraction
intensities arising from this class of systems may be used as
input to just the same holographic reconstruction algorithm as
developed for the DLEED case. Two important restrictions
for this type of application have to be mentioned: 
there must be only one elevated
adatom per surface unit cell and the unit cell must have a minimum size.
The first condition arises from the necessity of a unique
holographic reference wave as we will outline in the third section.
The second limit is based on the data density available and required. The approximate
minimum size of the unit cell could recently be estimated as a p(2$\times$2)
mesh \cite{Reuter98}. 
An upper size limit is drawn by experimental factors and the more likely appearance
of several adatoms per unit cell with increasing unit cell area.
A (7$\times$7) cell already appears to be too large as discussed in section III.

Still, a considerable number of ordered reconstructions remains open for a 
holographic analysis. 
In the present paper, our investigations are focused on the first 
successful application to an ordered and {\em a priori} unknown complex 
structure case, the example of the SiC(111)-(3$\times$3) superstructure. 
This surface phase is of considerable interest in
current crystal growth investigations \cite{Tanaka94} of the
promising semiconductor material SiC \cite{SiC-Band}. 
Previous STM work \cite{Kulakov96,Li96,Starke98_2} had revealed a 
single large protrusion per surface unit cell. 
So, this reconstruction seemed particularly suited for a first application 
of holographic LEED to ordered surfaces meeting both requirements outlined
above, i.e. sufficient unit cell size and the presence of a single elevated atom.
By comparing the holographic images obtained from experimental data and from
calculated intensities for fictitious models deviating from the real 
surface geometry we illuminate new aspects of the validity and 
possible shortcomings of the new method.

The paper is organized as follows: in the next section we recall the
holographic reconstruction algorithm using diffuse LEED intensities.
Thereafter we describe the relation between diffuse and discrete
intensities and give arguments under which circumstances the
holographic algorithm may readily be applied to conventional spot
intensities. This is followed by the reconstruction of an atomically
well resolved image from the experimental LEED intensities measured
for the SiC(111)-(3$\times$3) phase. Section V shows that the spatial
depth accessible by the method is rather large, which allows to determine
such important features as the stacking sequence of deeper layers.
In section VI, we illuminate the role of periodic vacancies within the unit
cell acting as additional holographic reference waves. Then we
address the issue of intensities arising from substrate relaxations
such as buckling and
finally discuss the use of the holographic information for a complete
surface structure analysis in the case of SiC(111)-(3$\times$3), 
whose precise real space structure is described in more detail elsewhere
\cite{Starke98_1,Schardt98}.

\section{The holographic reconstruction algorithm}

\begin{figure}
\epsfxsize=0.5\textwidth \epsfbox{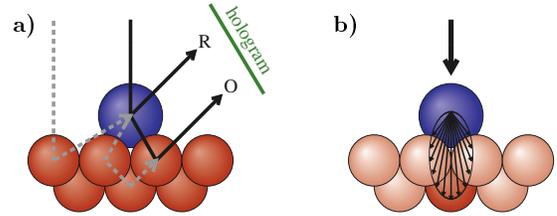}
\caption{Schematic display of the holographic interpretation of the adatom scattering:
a) electrons finally scattering at the adatom form the {\em reference} wave {\em R},
those subsequently hitting one substrate atom represent the kinematic {\em object} wave {\em O} 
(solid lines). The dashed lines display possible multiple scattering events providing
dynamic contributions to the reference and object wave (see text for details).
b) Pronounced forward scattering at the beam-splitter indicated by different length of the 
arrows in different directions.}
\end{figure}

The holographic approach in DLEED makes use of the fact, that all
measurable diffuse intensity outside the sharp substrate Bragg spots
necessarily has been caused by at least one scattering event at one of
the disordered adsorbates on top of the (unreconstructed) crystal:
scattering exclusively within an ideal bulk-terminated substrate can
only lead to diffraction intensities at Bragg spot positions. 
In that sense the adsorbate atom can be viewed as a prominent scatterer
which, acting as a beam-splitter, 
provides a natural separation of all scattering paths
as depicted by the solid lines in Fig.~1(a): electrons whose final
scattering is by an adsorbate form the {\em reference} wave $R({\bf k})$,
while those scattered subsequently by substrate atoms before reaching
the detector provide the {\em object} wave $O({\bf k})$ \cite{Saldin90} 
(where ${\bf k} \equiv ({\bf k}_{\|}, k_{\perp})$ is the wavevector
of the detected electron, with the components ${\bf k}_{\|}$ parallel to,
and $k_{\perp}$ perpendicular to the surface). This allows to interprete
the diffuse intensity as the interference pattern of these two contributions.
Hence, the local surrounding of the beam-splitter atom should be extractable
by a phased 2D Fourier transform of the data \cite{Barton88,Saldin90}.
However, the above interpretation does not include that multiple scattering
adds unwanted contributions to both reference and object wave as indicated
in Fig.~1(a) by the dashed lines. A considerable improvement in the image
quality could be achieved by combining several DLEED patterns measured at
different electron energies \cite{Barton91}. The corresponding multi-energy
reconstruction algorithms include a 3D integral transform and try to single
out the contributions due to the kinematic object wave, suppressing the
unwanted effects caused by the multiple scattering of the low energy electrons.

Yet, the pronounced forward scattering of the beam-splitter
led to only a selective appearance of the atoms in the reconstructed
local adsorption geometries, depending on whether they were located
within the forward scattering cone \cite{Wei92},
cf. Fig.~1(b). The implied necessity of combining several 
(at least two) data sets taken at different angles of incidence to
deduce the complete 3D surrounding of the beam-splitter \cite{Wei94}
could be overcome with the introduction of an improved reconstruction 
algorithm proposed by Saldin and Chen \cite{Saldin95}.

This {\em Compensated Object and Reference wave Reconstruction by an
Energy-dependent Cartesian Transform} (CORRECT) \cite{Saldin95}
allows the calculation
of the real space distribution around the adsorbate
$\left| B({\bf r}) \right|^2$ (where ${\bf r} \equiv ({\bf r}_{\|}, z)$
is a position vector relative to the origin at the adsorbate with
components ${\bf r}_{\|}$ parallel to, and $z$ perpendicular to the surface)
via the following expression:

\begin{eqnarray}
B({\bf r}) &=& \int\!\!\!\int_{{\bf k}_{\parallel}} \left[ \int_{k_{\perp}} 
K({\bf k}_{\parallel}, k_{\perp};{\bf r}) \chi({\bf k}_{\parallel}, k_{\perp})
e^{-(ikr-k_{\perp}z)} dk_{\perp} \right] \nonumber \\
& & e^{i{\bf k}_{\parallel}\cdot{\bf r}_{\parallel}} d^2{\bf k}_{\parallel}.
\label{correct}
\end{eqnarray}

Note, that in contrast to previous reconstruction algorithms that
performed the involved 3D integral in a polar coordinate system (angle
and energy), the data input is provided on a cartesian grid
$({\bf k}_{\|},k_{\perp})$, which will be of importance when discussing
the step towards ordered superstructures in the next section.

The transform does not operate directly on the measured intensities $H$,
but rather on a contrast-enhancing and normalizing function

\begin{equation}
\chi({\bf k}_{\parallel}, k_{\perp}) = \frac{H({\bf k}_{\parallel}, k_{\perp}) -
H_{av}({\bf k}_{\parallel})}{H_{av}({\bf k}_{\parallel})}
\label{chi}
\end{equation}

with 

\begin{equation}
H_{av}({\bf k}_{\parallel}) = \frac{\int H({\bf k}_{\parallel}, k_{\perp})
dk_{\perp}}{\int dk_{\perp}}.
\label{hav}
\end{equation}

It has been shown theoretically that the use of such a $\chi$-function
helps to partially remove the self-interference terms
$\left| R ({\bf k}) \right|^2$ and $\left| O ({\bf k}) \right|^2$ in
the DLEED intensity, which give rise to spurious high values of the
real space distribution $\left| B({\bf r}) \right|^2$ in the vicinity
of the origin \cite{Saldin95}. Additionally, $\chi$ has been designed
in such a way as to suppress modulations in the DLEED patterns
that arise from some partial ordering among the adsorbates \cite{Saldin97}.

The last part in the expression to be described is the integral kernel
which corrects for the anisotropy of the reference wave.
In a zeroth order approximation it can be written 

\begin{equation}
K({\bf k}_{\parallel}, k_{\perp}; {\bf r}) = \left[
\frac{f_a({\bf k}_i\cdot{\bf \hat{r}}) + C}{r} \right]^{-1}.
\label{kernel}
\end{equation}

\noindent
Here
$f_a({\bf k}_i\cdot{\bf \hat{r}})$ is the atomic scattering
factor of the adsorbate, ${\bf \hat{k}}_i$ the direction of electron
incidence, and $C$ the so called kernel constant (which we take to be real), 
and which represents an isotropic approximation to the backscattering by
the substrate prior to scattering by the adsorbate. Optimizing the value
of $C$ provides access to those atoms of the local adsorption geometry that lie
outside the forward scattering direction of the beam-splitter
\cite{Reuter97_1}. This allows the retrieval of the complete 3D
surrounding of the latter from data of normal incidence alone. The
algorithm in this present form has been shown to give reliable
images using theoretical \cite{Saldin95,Reuter97_1}, as well as
experimental DLEED data \cite{Saldin96,Saldin97}.

\section{Spot intensities versus diffuse distributions}

The original holographic reasoning \cite{Saldin90} was based on the
assumption that only one beam-splitting adsorbate atom
is present on the substrate surface. With several such adsorbates, each
time in the same local structure but without long range order
among them (lattice gas disorder), 
intensities simply add up in the low coverage limit leaving the resulting
diffuse distribution practically unchanged \cite{Starke96,Pendry84}. A
different situation emerges with the onset of order at higher
coverages and/or upon thermal annealing: additional modulations in
the DLEED pattern are created that eventually cause the breakdown of the
holographic algorithm \cite{Saldin97}.

For the case of a completely ordered superstructure of such adsorbates,
the modulations caused by the lattice factor concentrate 
the diffuse intensities to a series of
discrete superstructure - or fractional order  - spots when the unit mesh
of the adsorbate layer is larger than that of the
crystalline substrate. However, the simultaneous extinction of
diffuse intensity between the spots as caused by destructive
interference between the waves originating from different
adsorbate-substrate clusters, does not remove the crystallographic
information wanted: the energy dependence of the superstructure spot
intensities is the same as the one displayed by the corresponding
${\bf k}_{\|}$-positions in a diffuse distribution resulting from a
disordered adlayer in the equivalent local adsorption geometry
\cite{Heinz91}. The only restriction is that scattering between
such clusters has to be negligible, a condition satisfied even for
relatively small superstructures when using normal incidence data
\cite{Quasi,Mendez92}.

So, even though the perfect order among the adsorbates
significantly reduces the amount of available data, the few remaining
intensities are not masked by disturbing modulations as in the case
of partial disorder inhibiting the final ${\bf k}_{\|}$-integration
in equation (\ref{correct}). The superstructure spots can be thought of as
sampling the DLEED intensity distribution of the corresponding lattice
gas on a finite grid. Therefore, as suggested earlier \cite{Mendez92},
a DLEED holographic algorithm may in principle be applied to such
ordered superstructure systems with the only difference of a reduced
density of input data in ${\bf k}_{\|}$. This makes more apparent,
why the CORRECT algorithm is so particularly suited for the extension
to ordered phases: the data is provided on the appropriate cartesian
grid and only normal incidence is required. Interestingly, earlier
investigations on the information content of diffuse intensities
\cite{Heinz91}, as well as on the minimum data base of the algorithm
\cite{Reuter97_1}, showed, that the continuous diffraction distribution
resulting from disordered atomic adsorbates is already sufficiently
described when using a (3$\times$3) sampling grid. 
Information on a denser grid is largely redundant. Hence, there are
no drastic changes to be expected when making the transition from
disordered systems to phases with large superstructure cells like
the (3$\times$3) reconstruction of SiC(111) of the present paper.
This allows us to apply the algorithm developed for DLEED
without any modifications. However, it should be emphasized, that 
a further reduced data base in connection with superstructures
smaller than a (2$\times$2) can lead to aliasing effects in the Fourier-like
transform due to insufficient sampling \cite{Reuter98}.

The application of LEED holography to ordered surfaces involves several
practical advantages.
For the diffraction process it is irrelevant, whether the
beam-splitter is an externally adsorbed atom or intrinsically belongs to
the surface. Thus, besides ordered adsorption systems now also ordered 
substrate reconstructions can be investigated. 
Additionally, the measurement of discrete
spot intensities is much less delicate than that of the diffuse
intensities which are comparatively weak. 
The high signal-to-noise ratio of the bright spots allows easy subtraction
of contributions due to thermal diffuse scattering.
Also, at higher energies fractional spot intensities are not that much influenced 
by cross-talk from the bright substrate spots as is the case for
diffuse intensities \cite{Starke96,Mendez92}. Furthermore,
holographic LEED seems also suitable to tackle larger unit cell
reconstructions: the high number of fractional order spots generated in
these cases provides a fine sampling grid and ensures the proper working
of the integral transform. 
However, practical reasons also commend an upper limit for the unit cell size
as the increasing number of closely spaced
spots impairs a proper data acquisition, especially at
higher energies where more and more spots appear and weak spots are
disturbed by their bright neighbours.
A unit cell such as the (7$\times$7) on Si(111) \cite{Tong88,Takayanagi85} 
is probably already too large from an experimental
point of view as the accessible energy ranges become too small. 

In addition, it becomes more and more unlikely
that such a large unit cell contains only one elevated adatom
(the Si(111)-(7$\times$7) actually contains 12 adatoms).
This would violate the strongest restriction of the technique at its current
stage, i.e. the condition that only a single beam-splitter is allowed 
within each unit cell. Several such prominent atoms per unit cell would
lead to intermixing of their respective contributions as will be
demonstrated further below. This is all the more problematic, since the
actual number of elevated adatoms is just one of the quantities sought
in the structure analysis of an {\em a priori} unknown surface
(even though STM might help as in the present case). Future
efforts in methodologic improvements should hence be directed to overcome
the multiple beam-splitter problem, which did not occur in the previous
applications to simpler diffuse or ordered systems. As a consequence,
until there is a proper theoretical description of the detailed influences
on the reconstructed images, the systems to which holographic LEED is to
be applied have to be chosen with considerable care.

\section{Reconstruction using experimental data}

SiC is a material that displays most suitable electronic properties which
have made it a promising candidate for high power and high frequency devices.
Particularly, the (3$\times$3) phase of the SiC(111) surface has drawn
considerable interest in the last years, caused by the observed crystal
growth improvement \cite{Kong88}, that is achieved when this reconstruction
is stabilized under highly Si-rich conditions \cite{Tanaka94}. Its complexity,
which can already be deduced from an extensive debate in the literature
\cite{Kulakov96,Li96,Starke98_2,Kaplan89}, had hitherto prevented a
detailed structure analysis using trial-and-error methods. However, the high
number of fractional order spots caused by such a comparatively large surface
unit cell makes this phase an ideal candidate for a holographic investigation
in view of the reasoning outlined above.

The cubic 3C-SiC polytype was chosen, since its (111) oriented surface 
exposes only one definite stacking sequence \cite{Starke97}, i.e.
there is no coexistence of domains of different orientation,
which would have to be expected in the case of hexagonal polytypes
with different layer stackings possible at the surface \cite{Starke97},
and which would certainly complicate if not inhibit the interpretation of
the reconstructed images. Additionally, there is strong evidence from
comparison of experimental LEED intensities \cite{Schardt98}, as well
as from DFT test calculations \cite{Bechstedt97}, that the atomic
structure of the (3$\times$3) surface phase itself is rather independent
of the sample polytype. So, results obtained for 3C-SiC(111) can be
expected to hold also for other polytypes.

LEED I(V)-curves of the sharp diffraction pattern were measured 
in the energy range 50-300~eV 
using normal electron incidence. Details on the
data acquisition and sample preparation will be published elsewhere
\cite{Schardt98,Bernhardt98}. 
%It is worth mentioning that the restriction
%to normal incidence data, which can be well handled by the reconstruction
%algorithm, helps to considerably reduce experimental errors. The angle
%of incidence can easily be adjusted by comparison of symmetrically
%equivalent beams. The influence of a possible residual misalignment as
%well as of inhomogeneities of the fluorescent screen can be reduced 
%and the signal-to-noise ratio of the beams can then further be improved
%by subsequent averaging.
The low diffuse background and noise level allowed the recording of
14 fractional order beams closest to specular
reflection, which are symmetry-inequivalent at normal incidence.
Providing the measured intensities as input to expression (\ref{correct})
resulted in the 3D image displayed in Fig.~2: the real-space distribution
$\left| B({\bf r}) \right|^2$ is calculated on a grid of 0.2~{\AA}
resolution inside a cylinder of depth 6.0~{\AA} and a lateral radius
3.0~{\AA}, which is consistent with estimates on the lateral validity of
the algorithm \cite{Reuter98}. Small spheres are drawn at the grid points, 
indicating the reconstructed real-space intensity by their diameter 
which scales linearly with the intensity. As pointed out in previous
holographic investigations \cite{Saldin95,Saldin96,Saldin97,Reuter97_1},
this type of display permits a quick understanding of the essential
features of the structural unit determined holo\-graphically and will
therefore be used in all figures included in the present paper.

\begin{figure}
\epsfxsize=0.5\textwidth \epsfbox{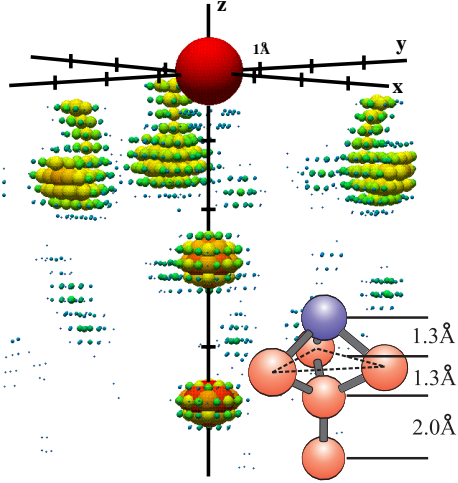}
\caption{Recovered local geometry of the SiC(111)-(3$\times$3) structure using experimental
data in the energy range 50-300~eV and kernel constant $C = 2.7$~{\AA}. The maximum noise
level in the image is 48~\% of the maxima denoting the atom positions (noise cut-off: 25~\%).
For details on the display procedure, see Section IV. The inset displays a schematic of
the retrieved adcluster geometry including chemical bonds and the approximate layer
distances as determined by holographic LEED.}
\end{figure}

The origin of the coordinate system is defined by the beam-splitter,
which is artificially added in the image as a black sphere 
to facilitate understanding. The highly Si-rich conditions under which
the (3$\times$3) phase is observed suggest this beam-splitter to be Si, the
scattering factor of which is consequently used for the computation of the
integral kernel (\ref{kernel}). However, the zeroth order approximation
of the latter is most sensitive only to the essential form of the atomic
scattering factor, which is very similar for most elements. Using carbon as
a beam-splitter in the computation consequently did not change the resulting
images considerably. The kernel constant $C$ in this expression is optimized
such that all atoms in the geometry appear with approximately equal
brightness \cite{Reuter97_1}. The highest disturbing intensity
at non-atomic positions (henceforth referred to as noise level)
is with 48~\% of the overall maximum value at an unprecedented low level.
This has to be attributed to the -- in comparison to the DLEED case --
much better quality of conventional LEED I(V) data and the increased
energy range available.

The image allows the unambiguous identification of the local adcluster
geometry formed by an adatom supporting trimer and two further atoms
vertically below the beam-splitter (see inset in Fig.~2). The rough layer
distances of 1.3~{\AA} (adatom-trimer), 1.3~{\AA} (trimer and first lower atom)
and 2.0~{\AA} (between lower atoms) correspond surprisingly well with
the (7$\times$7) DAS model of the Si(111) surface \cite{Takayanagi85}.
This already indicates that probably the complete
retrieved geometry corresponds to Si atoms on top of the SiC substrate.
Note, that the distorted form of the trimer atoms is an effect of the
scattering factor
in connection with the zeroth order approximation of this property
in the integral kernel (\ref{kernel}) \cite{Saldin95}. However, neither
the obtained spatial resolution, nor the exact position of the atoms inside
the geometry are the primary object or strength of the holographic analysis:
it is rather the direct and quick idea of a structural unit belonging to the
investigated surface.

The obtained tetramer formed by the adatom and the supporting trimer
is typical for hexagonal semiconductor surfaces. Its unambiguous
determination in the holographic image proves that only one of the
two possible orientations rotated by 60$^{\circ}$ with respect to each
other is present on the surface. This already excludes 
domains of differently rotated, i.e. coordinated clusters, 
as can also directly be deduced from the pronounced
threefold symmetry of the measured fractional order spot intensities
\cite{Schardt98,Bernhardt98}. It further rules out the model proposed 
first by Kaplan \cite{Kaplan89} of a (3$\times$3) mesh which in close
analogy to the (7$\times$7) DAS model \cite{Takayanagi85} contains
two such tetramers per surface unit cell, which in turn would
necessarily be differently oriented. It should be noted,
that this model was also inconsistent with STM investigations, which
clearly revealed only one elevated protrusion per unit cell, thus
strongly favouring models including a single tetramer 
\cite{Kulakov96,Li96,Starke98_2}. Since the atomic beam-splitter
has to be identified with the top adatom of this tetramer,
exactly these results ultimately enabled the application of LEED
holography to this structure: the obligatory uniqueness of the
beam-splitter excludes DAS-like models with two tetramers per surface
unit cell from the class of systems accessible under the current
state of theory.

We should recall now that we are dealing with an {\em a priori}
unknown structure. Although the low noise level in the image may appear
very convincing, it has to be recognized that the strong multiple
scattering combined with the anisotropic scattering factors for
low energy electrons may lead to serious artefacts in the images that
would not easily be distinguishable from real atoms. Since the 
multi-energy algorithms developed for holographic LEED can only
suppress, but not completely eliminate these effects, it is often
advisable to vary the used input energy range. In view of the fact
that scattering factor anomalies may sometimes even lead to an
increased image quality when reducing the number of included energies,
the stability of the obtained result under such variations can
significantly increase the confidence in the deduced local geometry.
Dividing the experimental data into various
subsets always resulted in equivalent images, which strongly confirms
the structural unit determined. In general, one has to admit that
the weak scattering power of light elements like Si and C might provide
a favourable case for holography as multiple scattering contributions
are expected to be smaller than for example for transition metal crystals. 
However, recent results for the system O/Ni(001)-p(2x2) \cite{Reuter98} 
make us believe, that the developed algorithm does also work for 
stronger scattering materials.
Now, although the experimental result appears convincing a test of the validity
and sensitivity of the method seems adequate, which is presented in the 
next sections using simulated intensities from various fictitious models.

\section{Vertical sensitivity and stacking of deeper layers}

The simplest model consistent with the atomic positions obtained from the
holographic reconstruction would be just an adatom on top of a SiC bilayer.
Yet, such a model appears improbable since it would not account for the
strong silicon enrichment at the surface as detected by earlier Electron
Energy Loss (EELS) and Auger Electron Spectroscopy (AES) results
\cite{Kaplan89}, which we discuss in detail elsewhere \cite{Schardt98}.
Assuming therefore the tetramer as the essential structural unit -- presumably
formed by Si atoms in view of the EELS and AES results --
the first question to be verified is its position on the underlying
substrate. The simplest possible solution would be to directly place it
somehow on top of the SiC sample, to which the two further atoms showing
up in the holographic reconstruction would then belong. Given the bilayer
stacking sequence of this material in the [111] direction it is, however,
impossible to find any location consistent with two atoms directly on
top of each other as predicted by Fig.~2. Even though the holographic
method is most reliable in just this direction vertically below the
beam-splitter, in which atoms show up already without the scattering factor
compensation by the integral kernel (\ref{kernel}), there is yet no definite
certainty on the limit of the algorithm's validity for deeper layers:
all previous investigations with DLEED data had dealt with rather simple test
structures, which were already completely determined by the atomic positions
in the first two layers. Hence, the now performed calculation of
$\left| B({\bf r}) \right|^2$ up to a depth of 6.0~{\AA} raises concerns
whether the lowest lying atom identified at 4.6~{\AA} might already be
outside such a limit.

\begin{figure}
\epsfxsize=0.5\textwidth \epsfbox{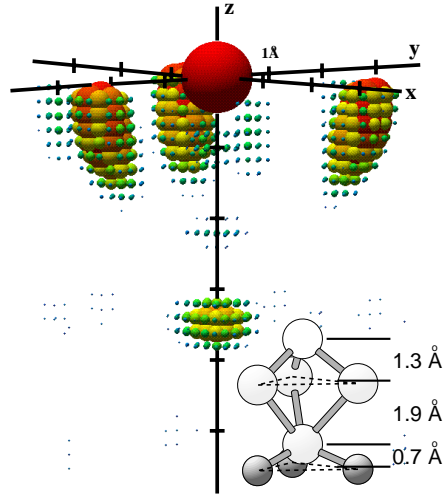}
\caption{Same as Fig.~2, but using simulated I(V) curves of a simplified Li/Tsong model
as described in section V. The electron energy range was 146-300~eV, the kernel constant
$C=5.0$~{\AA} and the maximum noise level is at 46~\% of the maxima at the atom positions
(noise cut-off: 25~\%).
The inset displays the atomic arrangement in the model assumed to calculate the intensities
used for the holographic reconstruction.}
\end{figure}

Consequently, for the moment we focus on only the tetramer and the
third layer atom of the holographic image. This arrangement 
suggests a cluster position
in which each of the three trimer atoms is located in a hollow site 
of the topmost hexagonal bilayer of the SiC substrate. Depending on
whether this position is fourfold coordinated, i.e. on top of a carbon
atom in the substrate bilayer (T4 site) or truely threefold coordinated
(H3 site), the tetramer would have to be oriented as in the SiC substrate
bilayers or rotated by 60$^{\circ}$, respectively. The adatom would then 
reside on top of a Si atom in the bilayer, which corresponds to the
third layer atom of the holographic image. No further atom would be present
2~{\AA} below the latter in either site geometry. Note, that this geometry corresponds
to the model proposed by Li and Tsong on the basis of their STM work
\cite{Li96}, which assumed such coordinated tetramers in a (3$\times$3)
periodicity directly on top of the SiC. 
In order to verify whether the atom additionally appearing 
in Fig.~2 is an algorithmic artefact or not,
we thus simulated theoretical LEED I(V) curves of an adcluster model for the
identical number of fractional order beams and the same energy range as
in the experiment (details of these calculations will be published elsewhere
\cite{Schardt98}).
In order to focus on the depth information available from LEED holography
we chose the fourfold coordinated trimer atom positions, yet artificially
expanded the distance between adcluster and substrate to push the third
layer atom to a position 3.2~{\AA} below the beam-splitter.
A schematic view of this geometry can be seen in the inset of Fig.~3.
The resulting holographic image is displayed in Fig.~3,
showing the expected tetramer unit plus the third layer atom, but also
consistently not indicating any sign of possible artefacts vertically below
the adatom. Only the carbon atoms of the substrate SiC bilayer are still within
the reconstruction volume, but do not appear in the reconstructed image (Fig.~3). 
However, one has to consider that in the geometry chosen
they are 4.3~{\AA} apart
from the beam-splitter and 1.9~{\AA} off the vertical axis, and in addition
represent comparatively weak scatterers. This distance -- not in 
forward scattering direction -- probably represents the detection limit
at least for a weak scatterer.
However, in view of the absence of artefacts in the holographic image obtained
for our test model, the presence of two atoms vertically below the adatom 
as indicated in the real space reconstruction from the experimental data has 
to be assumed correct.
Furthermore, the pronounced appearance of the lowest atom might indicate that
it is silicon since a weakly scattering C atom should not be detectable at
that depth.
Our test case resulting in the image shown in Fig.~3 thus clearly rules out 
the possibility of the Li/Tsong model,
whose bilayer stacking sequence results in fourth layer atoms off the
vertical axis and which is thus incompatible with the lowest
atom in the image obtained from the experimental data.

This further emphasizes the importance of the retrieval of the two deepest
atoms in the local geometry. Since only small deviations from the bulk positions
are usually to be expected in such deep layers, the location of each of
these atoms uniquely determines the complete stacking sequence of the 
corresponding entire layer. Hence, even though the obtained structural unit
itself may contain only a small number of atoms, its consistent embedding
into the surface unit cell subsequently can reveal a quite important further 
fraction of the investigated surface.

\section{Can a vacancy act as a beam-splitter?}

The confirmation of the lowest atom inside the revealed structural unit,
whose on top stacking is inconsistent with a SiC bilayer at the very
surface, necessitates to include an additional Si adlayer in the
crystallographic model. Such an adlayer
between tetramer and substrate had already been included in the original
DAS-like model, since the EELS and AES results indicated the strong
presence of Si-Si bonding in the highly Si-rich surface \cite{Kaplan89}.
In order to bring this otherwise very reasonable model in accordance
with the STM data described above, Kulakov {\em et al.} proposed the
absence of one of the two tetramers per surface unit cell \cite{Kulakov96}.
In the language of the DAS-type models the top atom of the trimer represents
an adatom on top of a Si bilayer. In the DAS-unit cell one of the two
adatoms is located on a piece of bilayer in faulted orientation
as indicated in Fig.~4(a). One would
expect an energetical difference between the adcluster which follows
the substrate stacking direction and the one, which introduces a
local stacking fault in the adlayer. Thus it is plausible that in the end
exclusively the more favourable type would be present on the surface.
Such a model, hence including only a single tetramer with definite
orientation in each (3$\times$3) unit cell, could explain not only
the single protrusion in the STM images, but also the threefold
symmetry of the LEED pattern. What remains, is the question
which of both orientations is actually realized.
Note, that the orientation of the adcluster can be deduced from
a comparison with a previous LEED analysis of the (1$\times$1) phase on 
the same sample \cite{Starke98_1,Starke97,Starke97_2}. 
However, we demonstrate a verification of this assignment using 
test calculations,
a method that could generally be applied in cases where no
independent analysis of the substrate is available.

\begin{figure}
\epsfxsize=0.5\textwidth \epsfbox{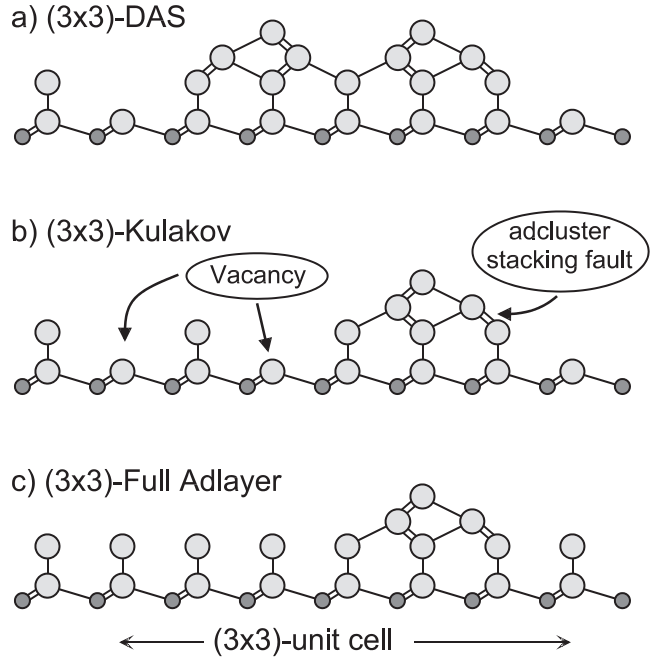}
\caption{Schematic side view of different models for the SiC(111)-(3$\times$3) reconstruction
in a projection parallel to the [1$\bar{1}$0]-plane 
(Si atoms are depicted by large spheres, C atoms by small darker spheres. 
Bonds within the projection plane are drawn as single lines, 
double lines represent two bonds pointing out of the projection plane by 
+60$^{\circ}$ and -60$^{\circ}$, respectively.)
a) DAS-model containing two adatoms and one cornerhole per unit cell
as proposed by Kaplan \cite{Kaplan89}.
b) Single adatom model containing two cornerholes with a local stacking
fault in the Si-bilayer fragment underneath the adatom; derived from
the model by Kulakov et al. without the stacking fault \cite{Kulakov96},
c) Single adatom-trimer-cluster residing on a complete monolayer in
(1$\times$1) periodicity with all cornerholes filled. The bilayer fragment
underneath the adatom again represents a local stacking fault.}
\end{figure}

Both Kulakov-type models, with the adcluster either introducing or not
introducing a local stacking fault, comply with the holographic image
from the experimental data, when identifying the upper of the bottom
two atoms as belonging to the Si adlayer and the other one as a Si
atom of the substrate's topmost bilayer. Yet, the subsequent
interplay between LEED holography and conventional LEED can do better
than that: when reconstructing images using theoretically simulated
data, the orientation of the substrate inside the given coordinate
system is known. Depending on the resulting orientation of the adcluster
in the image -- or to be exact, the tetramer of four atoms representing
the Si bilayer underneath the adatom --
when choosing one of the two equivalent beam assignments
in the LEED pattern, its orientation with respect to the bulk can 
be deduced even without seeing the latter in the reconstructed image
itself. From the result shown in Fig.~3, we therefore know the orientation
of an unrotated tetramer. Comparing this with the holographic image obtained
from the experimental data (cf. Fig.~2) we find that the adcluster geometry
involves a local stacking fault of the Si bilayer as shown in Fig.~4(b).
Hence, of all the previously existing models of the SiC(111)-(3$\times$3)
phase, LEED holography would only be fully consistent with the Kulakov
model in the local stacking fault version.

\begin{figure}
\epsfxsize=0.5\textwidth \epsfbox{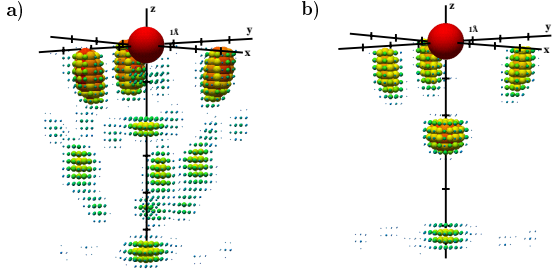}
\caption{Same as Fig.~2, but using simulated I(V) curves of a simplified Kulakov model
as described in section VI. The electron energy range was 146-300~eV and the kernel
constant $C=5.0$~{\AA}. a) geometry including cornerholes in the Si adlayer, 
cf. Fig.~4(b): maximum noise level at 76~\%, 
b) geometry with filled cornerholes in the Si adlayer, cf. Fig.~4(c): 
maximum noise level is at 29~\% (noise cut-off: 25~\%).}
\end{figure}

In order to further verify this conclusion, we
simulated LEED I(V) curves for this model, whereby, however, small
deviations from bulk-like positions as for example induced by
dimerization were not considered. Surprisingly, the corresponding
holographic image displayed in Fig.~5(a) is of considerably worse
quality than the previous results. Although all five atoms of the
expected structural unit show up, their overall configuration is
badly distorted and high noise in form of three concentrated artefacts
prevents the unambiguous distinction between real atoms and false
contributions. Since the image from ideal theoretical data should only
be better than the one from the experiment, the situation in the latter has
somehow to be more favourable for LEED holography than so far assumed,
which, of course, demands clarification.

As mentioned above, a severe source of an algorithmic breakdown at its
current level of development is given by the multiple beam-splitter
problem. Therefore, we reconsidered the atomic arrangement of the
Kulakov model under this point of view. Since it was derived in close
analogy to the DAS model, its Si adlayer is not completely closed but
contains vacancies to relax the stress induced by the lattice mismatch
\cite{Kulakov96}. These so called cornerholes appear just like in the
Si(111)-(7$\times$7) structure \cite{Takayanagi85} and break the (1$\times$1)
periodicity as much as the identified beam-splitting atom on top of
the tetramer. Consequently one might also concede them the {\em same}
holographic interpretation. 
In the sketch of the Kulakov model in Fig.~4(b) it can be seen that the
vacancy is even enlarged by the removal of one adcluster from the
DAS model, cf. Fig.~4(a).
It may appear difficult
at first glance to imagine a vacancy as a possible beam-splitter,
but speaking in terms of missing wave contributions to achieve the
destructive interference corresponding to a perfect (1$\times$1)
adlayer helps to understand its influence on the fractional order
beams. This would be equivalent to replacing
the vacancy by a pseudo-adatom with the same dynamic scattering
behaviour.

From this point of view, we would have to conclude that the Kulakov
model contains various distinct beam-splitters per surface unit cell,
whose respective wave contributions could consequently interfere and
completely prohibit a holographical interpretation. However, the strong
damping of the low energy electrons makes us hope, that the dominant
contribution in the reconstructed image is due to the most elevated,
periodicity breaking atom in the surface unit cell and that the existence
of further extra atoms or vacancies in the superstructure unit cell
leads "only" to image disturbances although they might be considerable. 
This interpretation
would help to understand why Fig.~5(a) basically shows the local environment
of the top tetramer atom plus artefacts and some distortions that may then be
due to the cornerhole vacancies. To test this line of thought, we 
simulated LEED I(V) data of exactly the same model as before, but filling
the vacancies with Si atoms at bulk-like positions in the Si adlayer
as shown in Fig.~4(c).
The resulting image in Fig.~5(b) is of impressive clarity and contains
all essential features of the result obtained with the experimental
data. We take this as a strong indication of the correctness of our reasoning,
although we want to stress again that there is yet no {\em proper}
theoretical treatment of the multiple beam-splitter difficulty in LEED
holography.
It should further be noted that in our test model, cf. Fig.~4(c) we only filled the
Si monolayer. The adatom supporting trimer atoms are still not repeated with 
the bulk periodicity and thus break the (1$\times$1) periodicity, too.

\section{Influence of substrate reconstructions}

What had started as a pure necessity to ensure the correct working of the
holographic algorithm, subsequently turned out to be the last required piece
for the solution of the (3$\times$3) puzzle. Since LEED holography seemed
only fully consistent with a Kulakov-derived model containing one tetramer
in local stacking fault orientation on a closed Si adlayer without
cornerholes, a careful reconsideration then showed that indeed there had
been no other reason for including the cornerholes at first hand but the
sole analogy to the DAS model. The situation for SiC(111)-(3$\times$3),
where the Si adlayer shows an intrinsic lattice mismatch of 20~\% with
respect to the underlying substrate, might however require a different
form of relaxing the lattice strain under simultaneous dangling bond
saturation than the cornerhole and dimerization principle underlying the
homoepitactic Si(111)-(7$\times$7). As a further hint, the obtained STM
images of the (3$\times$3) phase \cite{Starke98_2,Starke98_1} never
showed comparably strong cornerhole depressions as visible on the silicon
surface \cite{Becker85}.

Consequently, the thus most favoured model with filled cornerholes was
input to a refining LEED and Density Functional Theory (DFT) analysis.
Even though the holographic results had considerably reduced the
multi-parameter space for the trial-and-error search of both methods, it
should however be emphasized that the remaining structure determination
was still a tremendous task: there is a qualitative difference between a
coarse local beam-splitter surrounding that depicts a small fraction of the
huge surface unit cell and a detailed variation of all involved atomic
positions on a dense grid in steps of a few hundredths of {\AA}ngstr{\o}ms. 
It was in this
respect most gratifying that both analyses independently yielded the
same full (3$\times$3) structure by input of the holographically
recovered cluster: the resulting final {\em twist} model
\cite{Starke98_1} can indeed essentially be described as a SiC substrate
with a strongly buckled Si adlayer without any cornerholes plus one
tetramer per surface unit cell consisting of a trimer and one adatom.
These trimer atoms and the Si adlayer below locally resemble a Si bilayer
in stacking fault orientation. A more detailed
description of the exact model and both analyses involved
will be given elsewhere \cite{Schardt98,Furthmuller98}.

\begin{figure}
\epsfxsize=0.5\textwidth \epsfbox{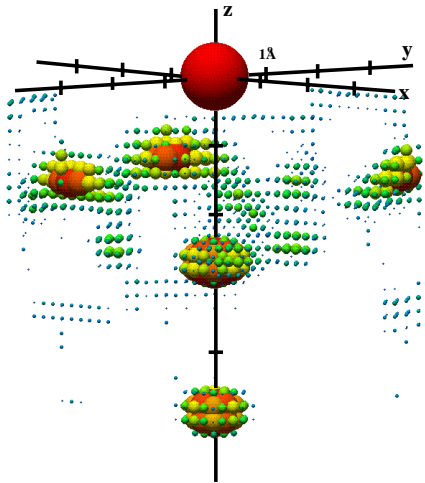}
\caption{Same as Fig.~2, but using simulated I(V) curves of the final twist model
as described in section VII. The electron energy range was 65-300~eV, the kernel
constant $C=0.75$~{\AA} and the maximum noise level is at 51~\% (noise cut-off: 25~\%).}
\end{figure}
 
Yet, we have to realize that the bond optimizing relaxations inside the
adlayer also contribute to the superstructure spot intensities and
might consequently affect the working of the holographic reconstruction
algorithm: each atom, that has left its (1$\times$1)-like position,
has in principle to be regarded as a possible additional beam-splitter in
view of the discussion in section VI. Therefore, as a final test, we used the
I(V) curves calculated for the exact geometry of the optimized model as input to the
holographic algorithm. The resulting image is displayed in Fig.~6 and
shows exactly the same local adatom environment as the experimental data,
which we - even in the presence of buckling -- take as a final proof of the
validity of our new method. In accordance with the experience
made recently with other, simpler structures \cite{Reuter98}, the effect
of the slight deviations from bulk-like positions on the reconstructed
image is apparently negligible and leads only to an increased overall
noise level, which can be seen comparing Fig.~5(b) and Fig.~6, whose
underlying structure differs exclusively by just these substrate
relaxations. All this becomes more understandable, when recalling,
that it is again only the {\em difference} in the outgoing
wavelets arising from buckled atoms that can act as a conduit for diffraction
intensity in the fractional order beams. The (additionally damped)
contributions due to these shifts can hence be regarded small with
respect to the major rupture of periodicity caused by the introduction
of a completely new and elevated atom such as the adatom in the present
structure. In this context, it is also important to notice, that the
majority of such shifts is far below the resolution capabilities
of LEED holography at its present stage, which can typically be stated
to be $\approx$ 0.6~{\AA}.

\section{Conclusion}

In the present paper we described in detail the contribution that
holographic LEED can provide, when applying it to a complex superstructure.
Using a holographic interpretation of fractional order spot 
intensities, a 3D image of the local geometry
around an elevated, periodicity breaking adatom can be retrieved.
The structural unit thus obtained has to be consistent with the real
space atomic structure and can be used to 
considerably reduce the multi-parameter space and possibly
enable the trial-and-error search of geometry optimizing methods like
quantitative LEED and DFT energy minimization. We exemplified this
for the case of the SiC(111)-(3$\times$3) phase, where the
holographically derived adcluster {\em directly} rules out the
majority of all previously existing models of this surface and
whose geometry has now been fully confirmed by the final {\em twist} model
obtained independently by conventional LEED and DFT. This application
additionally marks the first example that a holographic inversion of
LEED data actually played a crucial part in the determination of a
complex and {\em a priori} unknown structure.

We have illuminated the power and the limits of this new holographic LEED 
method for ordered surfaces.
Even in the best of all cases the obtained image is 
restricted to the local geometry around the prominent beam-splitter.
Only for very simple surfaces, this uniquely determines all atomic
positions inside the unit cell. Furthermore, exclusively for such simple
test surfaces has this still developing method been thoroughly tested
so far. There, some severe problems like the multiple beam-splitter
problem encountered in the present investigation do usually not arise
and have therefore not yet been theoretically treated. Consequently,
the systems, to which holographic LEED is to be applied, have for
the moment to be chosen with great care.

Nevertheless, regarding the immense problems that quantitative LEED
and DFT face with complex, large superstructure systems, every
directly obtainable pre-information is highly welcome. In this view,
the possibility of a holographic approach raises hopes that a new
ally in the structure analysis of these ordered phases has been found.
It is particularly its {\em direct} simplicity that already now renders
holographic LEED such an ideal supplement to its established
trial-and-error brethren, even though the young method has still a long
way to go.

\section*{Acknowledgements}

The authors are indebted to Prof. D.K. Saldin and Dr. P.L. de Andres for many
helpful discussions.
This work was supported by Deutsche Forschungsgemeinschaft (DFG) in particular
through SFB 292.

\end{document}